\documentclass[12pt,a4paper,pdflatex]{article}
\pdfoutput=1
\usepackage{amsmath}                                       
\usepackage{amsthm}                                        
\usepackage{amssymb}                                       
\usepackage[round, longnamesfirst]{natbib}                 
\usepackage{graphicx}                                      
\usepackage{epstopdf}
\usepackage{colortbl}                                          
\usepackage{setspace}                                      
\usepackage{rotating}                                      
\usepackage{enumitem}                                      
\usepackage{booktabs}                                      
\usepackage{multirow}                                      
\usepackage{tabularx}
\usepackage{color}
\usepackage{setspace}
\usepackage{fullpage}
\usepackage{hyperref}                                      

\begin{document}

\setlength{\parindent}{0pt}
\setlength{\parskip}{1.5ex plus 0.5ex minus 0.2ex}

\author{Daniel Kapp\thanks{Daniel Kapp is a Ph.D. candidate at the University of Paris 1-Panth\'{e}on-Sorbonne and at the Paris School of Economics (e-mail:\htmladdnormallink{Daniel.Kapp@malix.univ-paris1.fr}{mailto:Daniel.Kapp@malix.univ-paris1.fr})} \and Marco Vega\thanks{Marco Vega is Deputy Manager of the Research Divison at the Central Bank of Peru and Associate Professor at the Universidad Cat\'{o}lica del Peru (e-mail:\htmladdnormallink{marco.vega@bcrp.gob.pe}{mailto:marco.vega@bcrp.gob.pe})}}
\title{Real Output Costs of Financial Crises:\\ A Loss Distribution Approach\footnote{We are grateful to Fabrizio Orrego, Fabrizio Coricelli, participants at the Central Bank of Peru Annual Conference (\emph{Encuentro de Economistas} 2011), Centre of European Studies Annual International Conference, and research seminars at both the Central Bank of Peru and the Economics Department at PUCP for very helpful comments. All remaining errors are our responsibility.}}
\date{March 2012}



\addvspace{2.0 cm}

\maketitle

\begin{abstract}
We study cross-country GDP losses due to financial crises in terms of frequency (number of loss events per period) and severity (loss per occurrence). We perform the Loss Distribution Approach (LDA) to estimate a multi-country aggregate GDP loss probability density function and the percentiles associated to extreme events due to financial crises.

We find that output losses arising from financial crises are strongly heterogeneous and that currency crises lead to smaller output losses than debt and banking crises.

Extreme global financial crises episodes, occurring with a one percent probability every five years, lead to losses between 2.95\% and 4.54\% of world GDP.

\addvspace{2.5 cm}


\addvspace{0.2 cm}


\end{abstract}

\normalsize

\newpage

\section{Introduction}

Financial crises have played a quintessential role after the collapse of the Bretton Woods system of fixed exchange rates. Episodes like the Latin American debt crises in the 80's, the 1987 Black Monday, the 1992-1993 ERM crisis, the 1994-1995 Tequila crisis, the 1997-1998 South East Asian meltdown, the 1998-1999 Brazilian and Russian crisis, the 2000-2001 Turkish crisis, the 2001 Argentine crisis and the 2007-2009 global financial crisis all resemble disaster events, just like hurricanes or earthquakes.

Like catastrophic events, financial crises can be characterized by frequency and severity. In fact, analysis made in the financial crises literature often refers to terms such as frequency and severity \citep[see][]{Bordoetal2001}. The insurance and operational risk theory and practice offer toolkits to analyze frequency and severity of losses as well as aggregate losses due to catastrophic or operational risk events. Given that the losses to country economies from financial crises (in terms of GDP drop or forgone GDP) are similar to catastrophic losses, we can apply the loss distribution approach (LDA) familiar in the actuarial literature to analyze frequency and severity of losses and thus study rare events and their probabilities \citep[see][]{Panger2006,Shevchenko2011}. To the best of our knowledge, there is no attempt to quantify the frequency and severity of financial crises using the LDA.

The advantage of the LDA is that we can analyze frequency of events and severity in a separate fashion and then combine them to obtain a global loss probability density function. In the study of GDP losses arising from financial crisis, it is of interest to study frequency and severity on their own as shown in \citet{Bordoetal2001}.

The LDA allows us to estimate a multi-country aggregate GDP loss distribution and thus estimate conditional losses in the event of a financial crisis occurring in the near future. We can also determine the probability of rare economic disasters as defined for example in \citet{Barro2006}. In contrast to \citet{Barro2006} however, we do not make any assumptions on the channel through which crises occur (e.g. catastrophic events like earthquakes or regular/cyclical disaster events).

In terms of methodology, we use the financial crises database of \citet{Laeveneta2008} to date financial crises across 170 countries from 1970 onwards. The number of such events over a predetermined period is called the frequency of events. We then estimate output losses per financial crises event with a number of methods. Afterwards, we aggregate country output losses across events over the period. Since a particular crisis event can generate output losses over various years, we set the span of analysis to be five years. Finally, we compound frequency and severity to generate a loss probability density of aggregate losses that allows us to report standard risk measures.

We find that output losses after financial crises are strongly heterogeneous and a large number of countries never recovered their pre-crises growth rates or trends. Also, we show that currency crises lead to smaller output losses than debt and banking crises, while the largest losses are found after debt crises. The presence of a debt crisis also exacerbates any of the other two forms of crises, while the presence of a currency crisis in the wake of a debt or banking crisis diminishes output losses through faster recovery. Banking and debt crises alone are found to be more severe than twin crises consisting of banking and currency crises or debt and currency crises.

The LDA approach leads us to conclude that mean worldwide costs of financial crises within periods of 5 years are in the range of 0.52\% to 0.81\% of 2005 world GDP. Extreme crises episodes, occurring with a one percent probability, can lead to losses between 2.95\% and 4.45\% of world GDP.

The analysis of losses produced in the paper can be a useful tool in discussions about the existence of insurance against the risk of financial crises at the aggregate level. For example, \citet{Caballero2003} proposes such an arrangement for emerging market economies.

In what follows, we will provide a short literature review and discuss the possibilities at hand to calculate output costs of financial crises. In section 3, we introduce the methodology of crisis identification and loss calculation, while the Loss Distribution Approach is explained in Section 4. Results are presented in Section 5. Section 6 discusses a potential form of international insurance and Section 7 concludes.

\section{The costs of financial crises }

In order to quantify the costs of financial crises, we need to define a metric. Studies on this subject use a varied set of cost measures. Costs have been estimated as fiscal costs, costs to the stock market, and output costs. Among studies analyzing the same kinds of losses, methodologies also differ. In addition, one of the main obstacles to measuring losses caused by financial crises is how to identify losses only due to financial crisis and not arising from other contemporaneous factors.


\subsection{Fiscal costs, costs to the stock market and output losses}

In an attempt to quantify the costs of banking crises to the economy, \citet{Hoggarth2002} consider direct resolution costs as well as broader welfare costs to the economy, approximated by output losses. They argue that resolution costs are a rather limited proxy for costs incurred through banking crises, as they may reflect a transfer of income from taxpayers to banks rather than costs imposed to the economy as a whole. The authors reason that there could be a positive correlation between fiscal costs and output losses if crises are systemic. On the other hand, if fiscal costs are a good proxy for effective crisis resolution, higher spending on crisis resolution should lead to lower output losses during a crisis period. No clear statistical relationship between fiscal costs and length of crises is found, while output losses and the length of crises do depict a clear positive correlation.

\citet{Frydl1999} presents a comparative analysis of prior banking crises studies. As one of the reasons for the non-significant statistical relation between resolution costs and crises length, he claims that resolution costs usually measure fiscal costs of banking crises, which are often subject to various errors and do not incorporate many indirect costs to the government or the economy. Having dismissed fiscal costs as a reliable indicator for crisis severity, \citet{Boydeta2000}, use the discounted value of corporate returns to measure the impact of crises. Under the condition that corporate profits represent a relatively constant fraction of total output, a decline in the real values of stock prices at the onset of a crisis in percentage terms is approximately equal to the decline in the present discounted value of total output.

With regards to the impact and depth of currency crises, possible measures to be considered (in addition to output losses) are the loss of international reserves and the depreciation of the real exchange rate \citep{KaminskyReinhart1999}. However, the most popular method is to proxy costs to the economy with GDP losses, given that economic growth is a natural final performance indicator. This is the approach we follow in this paper.

We identify two strains of literature approximating real GDP losses due to financial crises. The first uses a dummy variable approach to estimate growth losses over samples of countries, studies like \citet{Demirguc-Kunt2006}, \citet{Guptaeta2007}, \citet{HannaHuang2002} and \citet{Barro2001} follow this path. The second approach proxies welfare losses by comparing GDP during a crisis period with some estimate of potential output. \citet{Hoggarth2002} is one representative study of this latter approach. It estimates potential output assuming that output would have grown at the same constant rate based on its past performance. Various studies, such as \citet{Bordoetal2001}, \citet{Azizeta2000}, \citet{Frydl1999}, \citet{Boydeta2000,Boydeta2005} and \citet{Cecchetteta2009} calculate output losses from banking crises in a similar fashion, even though their trend estimates are based on differing pre-crises windows, methods, definitions about onsets, ends and durations of crisis episodes.

A main criticism of the dummy variable approach is that it can only identify average magnitudes of growth contractions associated with crises for all countries. It therefore does not seem to be well suited for highly heterogeneous losses. Output costs calculated through cross-section or panel data regressions are usually found to be lower than losses calculated based on output gap estimations.

In a comparative analysis, \citet{Angkinand2008} suggests that the output gap approach is more appropriate than the dummy variable approach in capturing the output costs of crises. This is so because the individual output costs across crises vary substantially. Our loss estimation methodology therefore follows the output gap approach.

\subsection{Output gap calculations}
\textbf{Determining the length of a crisis}

The literature does not offer a unanimously agreed method to date the beginning of financial crises. Banking crisis start dates are usually defined through a mixture of quantitative and qualitative criteria. \citet{Caprioeta1996} rely on the assessment of finance professionals. Including and expanding on this approach, \citet{Demirguc-Kunt1998} compile five studies of banking crises' starting dates. Other studies identifying banking crises' dates are \citet{Dziobeketa1997}, \citet{KaminskyReinhart1999}, and \citet{LindgrenGarciaSaal1996}.

The onset of a currency crisis is generally defined as a situation whereby a sufficiently large depreciation of the domestic currency occurs\footnote{Sometimes an additional criterion of increased speed of depreciation as compared to some prior time window is introduced.}, often accompanied by a loss in international reserves, while a debt crisis takes place in the event of a country defaulting or renegotiating all or parts of its private debt. The most recent financial crises compilation, widely used in previous empirical studies, stems from \citet{Laeveneta2008} and comprises banking, currency, and debt crises over the period 1970 to 2008.

To date the end of a crisis episode, one possibility is to define the end date based on ``expert'' opinions or on the ``consensus'' view from various studies. An alternative is to define the end of a crisis endogenously once a country returns to a certain pre-crisis \textit{growth rate} or recovers its potential output \textit{growth path}.

Studies determining the end of a crisis based on the recovery of the average \textit{growth rate} of a pre-crisis window are, among others, \citet{Bordoetal2001} and \citet{Azizeta2000}. Authors such as \citet{Boydeta2005} argue that summing up deviations from an estimated trend up to the point at which the observed \textit{growth rate} returns to its pre-crisis average is problematic since output typically remains well below its pre-crisis absolute output trend once the growth rate has recovered.

\citet{Cecchetteta2009} avoid calculating a counterfactual and define the end of a crisis as the point in time when real GDP has reached its absolute pre-crisis level. This method is problematic in at least two ways. First, the method does not take opportunity costs of foregone output growth into account. Second, the method implies that a crisis is only counted as such if output growth actually turns negative during the crisis year. It can be argued, however, that a financial crisis has negative effects without having caused an actual recession, e.g. through a transitory or permanent slowdown in growth. Moreover, since potential growth rates vary across countries, dating the end of a recession by reaching its pre-crisis level of real GDP can lead to an underestimation of total losses incurred.

\textbf{Estimation of a counterfactual}

To be able to measure output losses during crisis periods according to the methods described above, it is necessary to compare actual output with its trend level. There is a handful of approaches to estimate trend GDP levels. They differ mainly by the pre-crisis time window chosen, which in turn depends on the assumption about financial crises either following economic booms or a slowdown in economic activity.

\citet{Hoggarth2002} assume that output would have grown at a constant rate based on past growth performance and extrapolate linear three and ten-year trends, while \citet{Bordoetal2001} use five-year pre-crisis trends. Instead, \citet{Frydl1999} apply ten year pre-crisis periods and \citet{Boydeta2000} extrapolate linear pre-crisis growth trends.


A large part of the heterogeneity in the magnitude of output losses in the study of crises stems from the calculation of trend output. Studies like \citet{Kindleberger1978}, \citet{Borioetal1994} and \citet{Logan2001} find that banking crises follow economic booms. In this case, a trend estimated over a short period prior to a crisis would overestimate potential output and lead to an overestimation of crisis length and depth. On the other hand, studies like \citet{KaminskyReinhart1999} or \citet{Gorton1988} find that banking crises are often preceded by a slowdown of economic activity, in which case losses would be understated.

If one assumes that pre-crisis growth deviates (in either direction) from the long-term potential output growth path, one option is to increase the pre-crisis trend calculation period in order to capture mostly ``normal'' years. An alternative is to exclude a certain period prior to the onset of a crisis. Last, the Hodrick-Prescot (HP) filter can be applied to diminish the influence of booms or recessions on the potential growth path. While \citet{Azizeta2000} base potential output on the average output of the three years prior to crises, \citet{Bordoetal2001} use five-year pre-crisis growth rates and \citet{Hoggarth2002} calculate potential output trends based on ten-, three-, and one-year pre-crises growth rates; for a comparative study see \citet{Angkinand2008}.

Once we establish the counterfactual, we can estimate total output losses by adding up the difference between actual and potential output over the duration of the respective crisis.

Even though the general concept is agreed upon across the studies mentioned above, several methodological issues remain debated. Identification of crises accompanied by output losses varies among studies. While some authors include a crisis if output is below its trend or if output growth is negative during the crisis year, other studies include crises even though output is above its trend in the crisis year, given that output is below trend in the subsequent year \citep{Angkinand2008}. Further issues arise in the case of multiple crises per country within short periods of time. In the case that output has not yet recovered from one crisis at the point of outbreak of a following crisis, some studies choose to sum losses of subsequent crises and report a single loss, while others divide losses across crises or simply choose to exclude countries with multiple crises during the sample period.

\subsection{Identification of causality}
To assess the direction of causality between economic growth and banking crises, \citet{Hoggarth2002} compare a sample of 29 countries experiencing banking crises with neighboring countries which did not face banking crises at the same time. The hypothesis is that ``the movement in output relative to trend during the crisis period would have been, in the absence of a banking crisis, the same or similar to the movement in the pairing country'' \citep{Hoggarth2002}. Their analysis hints at the point that output losses are in most cases caused by banking crises and come as unforeseen events.

\citet{Bordoetal2001} find, across all countries and crisis periods considered, that recessions with crises are more severe than recessions without them. These results are in line with various studies such as \citet{Frydl1999}.

\section{Methodology}

\subsection{Crisis identification}

We use the financial crises dataset of \citet{Laeveneta2008}. Currency crises, banking crises, and debt crises are identified over the period 1970 to 2008. \citet{Laeveneta2008} identify banking crises on the basis of a number of quantitative and subjective criteria, such as a large number of defaults and a high quantity of non-performing loans. The starting year of a currency crisis is identified by building on an approach developed in \citet{FrankelRose1996}. Sovereign debt crises are reported in the case of sovereign defaults to private lending as well as in a year of debt rescheduling. The number of currency crises peaked during the early 1980's and the early 1990's with around 30 currency crises per year, while banking crises have in general been less frequent and peaked during the early nineties. The number of debt crises per year has been decreasing since the mid-1980's and debt crises have nearly ceased to exist until recently, see \citet{BKM2011}.

We identify the starting date of a financial crisis as the year of outbreak of any one of the three types of crises. Real GDP data is taken from the World Economic Outlook database and spans the time period from 1960 to 2010.

To calculate output losses caused by crises, the first step is to define whether a crisis has an impact on the economy. In the case where output is compared to a counterfactual, we identify a crisis if output in the crisis year is below its trend  level (trend output estimations are discussed below). In an alternative calculation of crisis losses, no counterfactual is established and a crisis accompanied by output losses is considered as such if output growth is negative during the crisis year.

For countries with multiple crises during the sample period it is possible that a crisis occurs before the economy has recovered from a previous crisis. In this case, we assign subsequent losses to the later crisis date, establishing a new counterfactual. This method is problematic, though other alternatives suffer from larger errors. Allocating output losses from subsequent crises to the first crisis would largely overstate output losses in various cases.

\subsection{Output losses}

We estimate three kinds of potential output trends and propose several cutoff points to determine the end of a crisis. In short, there is no perfect method to estimate an objective output trend. Every method presented shows both advantages and disadvantages depending on the assumptions about the mechanics of financial crises.

Following the literature, we estimate potential GDP after the onset of a crisis in three ways. We estimate an HP trend, where potential output during a crisis episode is based on the average HP growth rate of the ten and three year pre-crisis periods. In addition, we use average growth rates from three year and ten year pre-crisis time windows. We compare the losses against trend output levels to absolute losses (for episodes with negative growth) without considering opportunity costs. In total we have thirteen possible ways to measure output losses as well as their severity and frequency.

\begin{figure}[h!]
\begin{center}
\caption{Ecuador 1998}\label{graph1}
\includegraphics[scale=0.8]{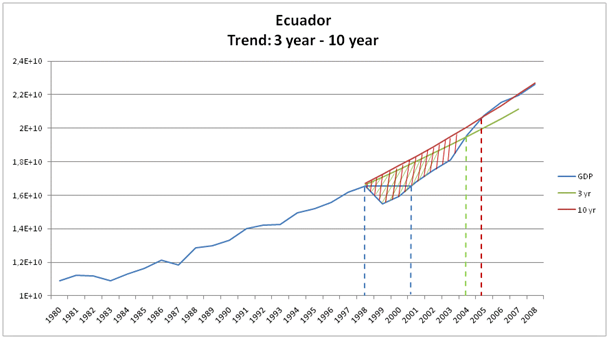}\\
\end{center}
\end{figure}

Output losses are calculated as the difference between actual real GDP and its trend level. A graphical example is given in Figure \ref{graph1}, depicting output during the Ecuadorian banking crisis, which shows that losses depend not only on the definition of the counterfactual but largely on establishing an end-point of a crisis.

According to the time output needs to recover its pre-crisis \emph{level}, the effects of the crisis lasted two years and led to an output loss of 10\% of GDP. If crisis length is calculated until real GDP growth reached its pre-crisis \emph{growth rate}, output losses occurred over a period of two and three years and led to output losses of 10.1\% and 22.4\% of GDP, calculated against a three and ten-year pre-crisis growth trend respectively.

\begin{figure}[h!]
\begin{center}
\caption{Ecuador 1998}\label{graph2}
\includegraphics[scale=0.8]{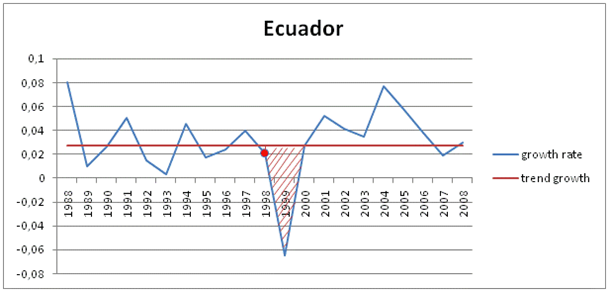}\\
\end{center}
\end{figure}

As shown in Figures \ref{graph1} and \ref{graph2}, this calculation most likely still does not account for the total output loss caused by the 1998 Ecuadorian banking crisis. The period of output loss increases to six and seven years, and output losses accumulate to 38.1\% and 51.4\% of GDP respectively  if losses are estimated until the level of real GDP recovers its three and ten-year pre-crisis \emph{trend}. As illustrated in Figures \ref{graph1} and \ref{graph2}, output losses seem to be underestimated if crisis recovery is defined as being completed at the point where the pre-crisis \emph{growth rate} or the pre-crisis \emph{level} of real GDP are recovered.

As mentioned above, we estimate losses using various trends and various cut-off points to determine the end of a crisis. In total, 13 loss estimations are presented.

We distinguish three definitions of recovery from a crisis. According to the first definition, a crisis ends once real output has reached the level of its counterfactual. The alternative is that recovery is completed once the average pre-crisis growth rate is resumed. As some countries never recover according to these definitions, accumulated losses against the counterfactuals based on linear three and ten year trends, based on simple averages of pre-crises growth, are considered over maximum periods of five and ten years, while the losses against trends based on the HP filter are allowed to accumulate over periods of maximum ten years. Note that an alternative potential remedy that we do not pursue here in cases where we have an infinite sum of losses is to perform present-value GDP losses by discounting future losses appropriately. In the absence of a counterfactual, a crisis is supposed to be ended once output reaches its absolute pre-crisis level of real GDP.

\section{The Loss Distribution Approach}

The estimated output losses across countries obtained in the previous section allow us to study the frequency and severity of losses. In the analysis of a financial crisis hitting the world economy, two usual questions appear: a) what is the frequency of financial crisis? and b) given a financial crisis, how severe is it? The frequency of financial crises is the number of such events over a specific period of time. Since a financial crisis duration spans generally more than a year, we choose a five-year reference period. The severity of a financial crisis is the amount of output loss incurred in each crisis episode.

Given the frequency and severity of losses occurring within 5-year periods, we can use the LDA common in the insurance and operational risks literature. \citet{Chernobaietal2007} and \citet{Cruz2004} provide detailed descriptions of the workings of the LDA. We estimate two possible parametric distribution functions commonly used to describe the frequency of events $n_t$ over a period of time $t$. We also estimate a set of six severity probability density functions\footnote{The set is composed of the gamma, exponential, generalized extreme value, generalized pareto, log normal and weibull density functions.} for events $z_{t,i}$. As opposed to the standard one year period, our $t$ represents periods of 5 years. This is because GDP losses due to financial crises consider losses over more than one year. The index $i$ tracks each event within the period of analysis $t$.

During a 5-year period, global losses are given by the sum of each loss event $i$ across countries in the sample.

\begin{equation}
  S_t=\sum _{i=1}^{n_{t}} z_{t,i}
\end{equation}

The random variable $n_t$ takes discrete values while the variables $z_{t,i}$ are non-negative (positive losses), real valued quantities. The aggregated loss $S$ depends on the realization of the discrete random variable ($n$) and the continuous random variable ($z_i$). Therefore, the aggregation $S$ is itself a random variable whose distribution has to be determined by convolution methods \citep{Panger2006,Shevchenko2011}.

Specifically, the frequency of loss events ($n$) has a probability distribution denoted by $p_n = Pr(N = n)$ while the loss severity $z$ has a density distribution and cumulative distribution functions denoted by $f_z$ and $F_z$, respectively. According to \citet{Panger2006}, the cumulative density function of $S$ is defined as:

\begin{equation}
\begin{array}{rcl}
  F(S) &= &Pr(\omega \leq S)  \\
       &  & \\
       &= & \sum_{n=0}^\infty p_n Pr(\omega | N=n)\\
       &  & \\
       &= &\sum_{n=0}^\infty p_n F_Z^n(S)
\end{array}
\end{equation}

Where $F_Z^n(S)$ is the n-fold convolution of the cumulative density function of $S$. A simple way to estimate $F(s)$ is via Monte Carlo simulations. First we draw $n$ from $p_n$ and then we draw $z$ from $f_z$ as many times as indicated by $n$. Last, we sum up the $z$ draws.

\section{Results}

In total, we observe 62 debt crises, 122 banking crises, and 196 currency crises. As some of these crises in effect form twin crises, we examine a total of 340 crises episodes. Depending on the method applied, we find between 110 and 219 contractionary crises episodes. While around 210 contractionary crises episodes are identified against HP trends, only 110 crises were accompanied by negative growth rates during the crisis year.

Average losses are higher compared to previous studies, this is so because econometric studies usually estimate losses over a sample comprising recessionary crises as well as crisis followed by a rebound in GDP, while we only consider crises leading to output losses (against a trend or absolute). From a total of 196 currency crises in our sample, between 90 and 120 crises (depending on the calculation method applied) have led to a loss of output compared to some measure of potential output, while 63 crises were accompanied by negative output growth. This result is in line with previous studies such as \citet{Guptaeta2007}, who find that about 60 percent of currency crises lead to output contractions while the rest were accompanied by output expansions.

The analysis that follows concentrates on a benchmark loss classification group that relies on the HP filter: losses until recovering average 10 year HP filtered GDP growth rates (HP10perc), the level of a 10 year HP filtered GDP trend (HP10trend), 3 year HP filtered GDP growth rates (HP3perc), and the level of a 3 year HP filtered GDP trend (HP3trend).

\subsection{The frequency of financial crises}

In order to carry out the LDA, we first estimate distribution functions for the frequency of losses from financial crises. We assume two commonly used distributions such as the Poisson and the Negative Binomial Distribution. The key parameter in the Poisson Distribution is $\lambda$ which is also the mean and variance of the number of losses. This is the drawback of the Poisson fit, the data in our benchmark case (the four HP trend counterfactuals) have numbers of crisis events with variances of about sixteen times the mean.\footnote{The number of crisis events for the other loss classifications have a variance to mean ratio of more than 13, except for the case where opportunity costs are not considered (ABS).} Therefore, the two parameter Negative Binomial distribution is a more flexible way to accommodate our data.

\begin{figure}[h!]
\centering
\caption{Estimated distributions for the frequency of losses}\label{graph3}
\includegraphics[scale=0.7]{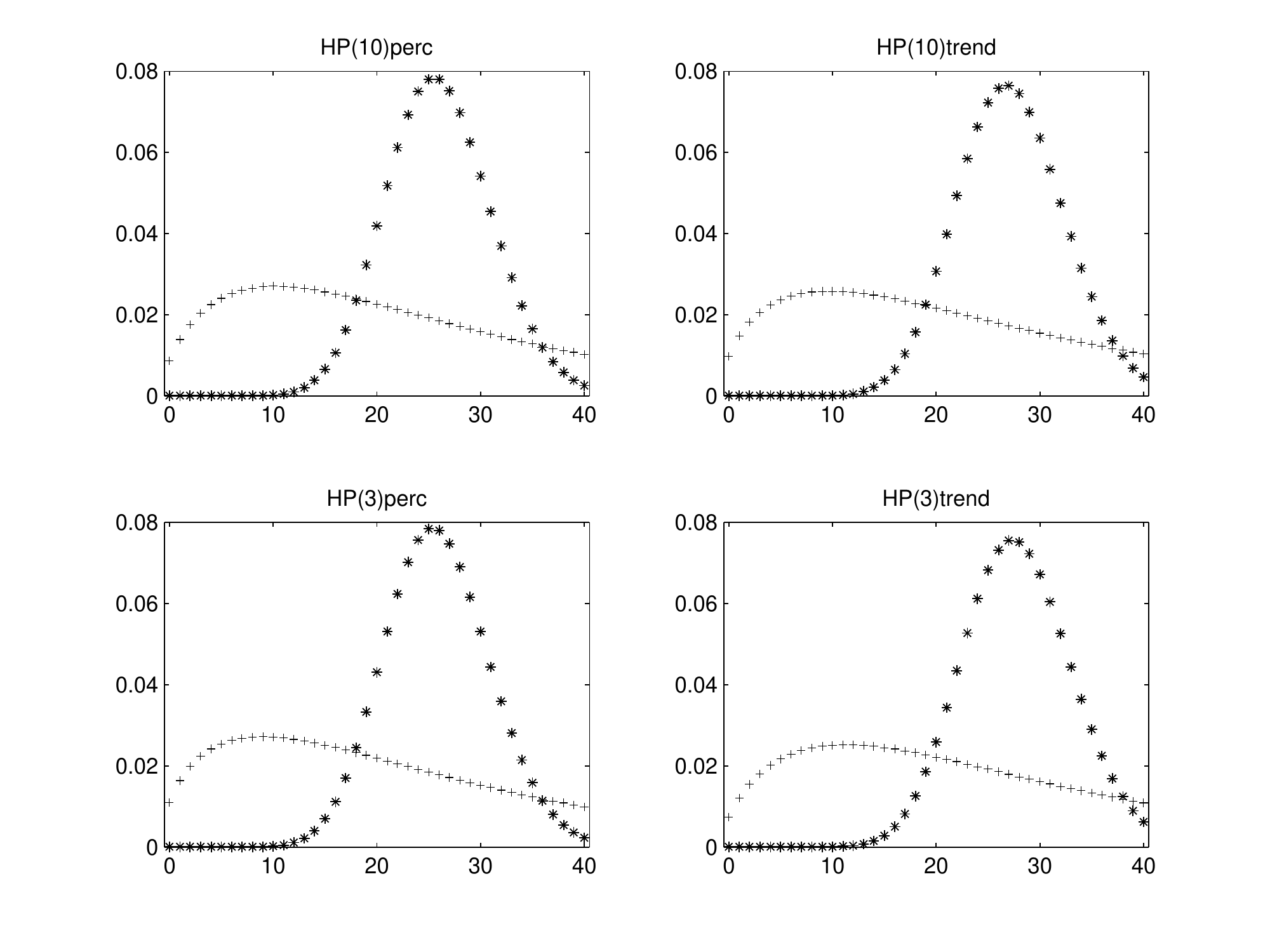}\\
\footnotesize{Note: +++: Negative Binomial, ***: Poisson. Horizontal axis measures the number of financial crisis in the world over a typical five-year period}
\end{figure}

In Figure \ref{graph3} we depict the estimated distributions for our benchmark loss classification. The Negative Binomial distribution has a lower mode but allows a more extreme number of losses relative to the Poisson distribution. The probability that the data come from the Negative Binomial is higher for all cases (the benchmark and all types of loss classifications). In all cases, the mean number of crises occurrence over five-year periods ranges from 20 to 30.

\subsection{The severity of financial crises}

The average accumulated loss caused by financial crises varies from 9\% of real GDP to 15\% of real GDP if output losses are accumulated against trends based on HP filtered data (Table \ref{table1}). In total, depending on the loss measure applied, 186 to 219 crises episodes are observed. Average losses are the largest when calculated against a ten year trend and if losses are considered until the level of trend output has been recovered over a maximum time span of 10 years.

Median losses lie between 4.9\% and 7.15\% of initial GDP and the most severe crises destroy up to three years of economic output.\footnote{Crises' percentage losses are calculated as the sum of the difference between real output observed and potential output after the onset of a crisis until its end, divided by the real output in the year of crisis onset. Losses are usually larger if calculated on the basis of other trend estimations as presented in the appendix.}

As can be seen in Tables \ref{table7} through \ref{table14} in \ref{append2}, alternative output losses calculated are very heterogeneous and large average percentage losses are driven by few especially severe crises events. Of these most severe events, potential output has not been reached again within a period of ten years. This result is in line with \citet{Furcerietal2011}, who find that the growth rate after debt crises eight years after the onset is still suppressed be nearly 10 percentage points. Amongst the largest losses observed are those of several Asian countries, namely Indonesia in 1998 and Thailand in 1997, both experiencing severe losses in the wake of the Asian crisis.

\begin{figure}[h!]
\centering
\caption{Mean excess over threshold for severity data. Both axes are in log scale of losses}\label{graph4}
\includegraphics[scale=0.8]{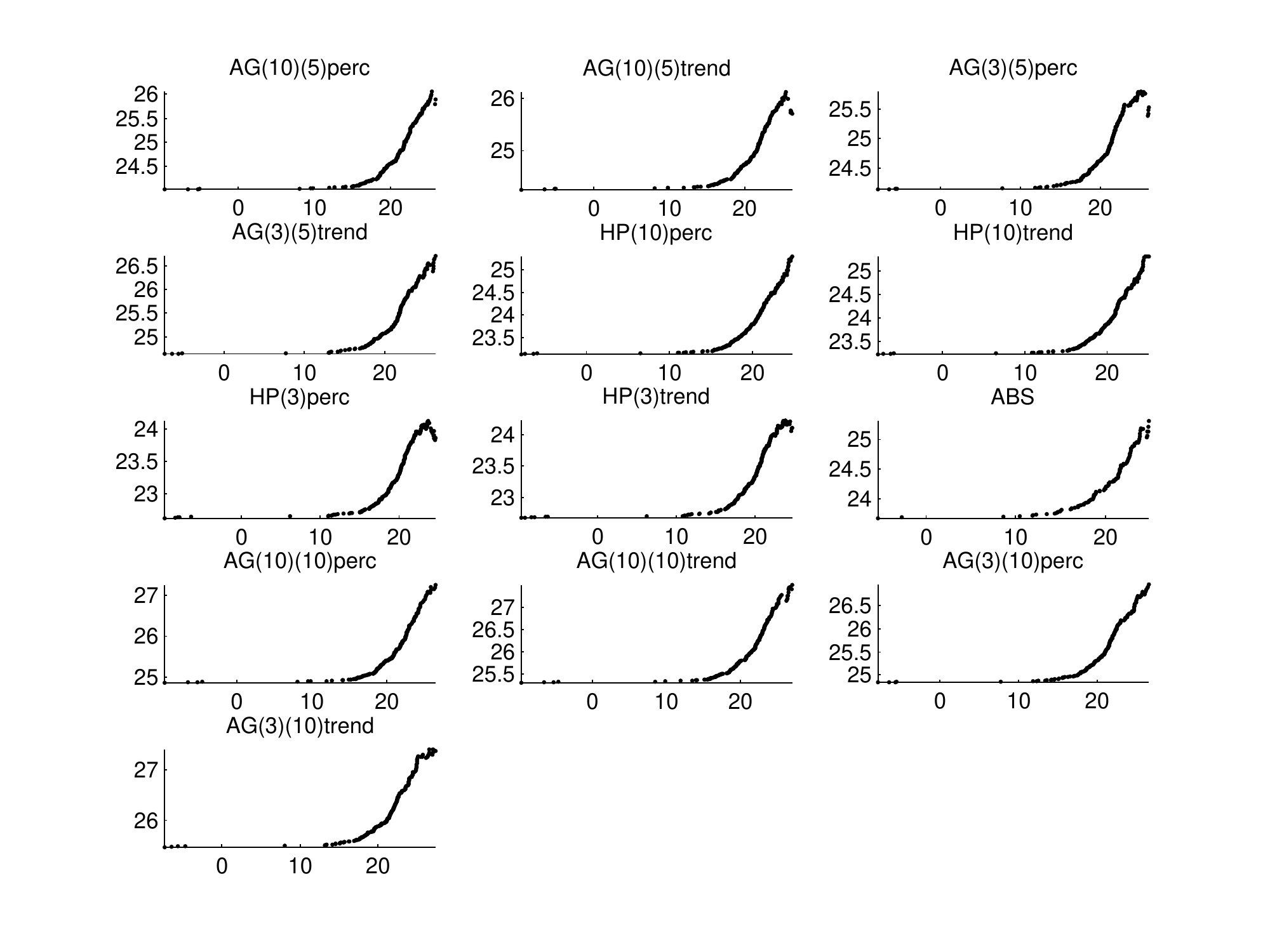}
\end{figure}

Severity data due to financial crisis have a key feature, it is extreme valued. In figure \ref{graph4} we see that the mean-excess-over-threshold plots\footnote{The sample mean excess plot is defined by: $mep = \frac{\sum_i ^n (X_i-u)I_{(X_i>u)}}{\sum_i ^n I_{(X_i>u)}}$, with $u>0$ and $I_{(.)}$ and indicator function.} have positive slope at the right extreme of losses. Figure \ref{graph4} shows that the long-tailed nature of severity data is generated by specific types of crises. Currency and twin currency-banking crises produce the more extreme type of losses.

\newpage
We fit the severity data with six possible probability density functions using the maximum likelihood estimator for the corresponding parameters. The six distributions are Gamma, Exponential, Generalized Extreme Value (GEV), Generalized Pareto, Log-normal, and Weibull. Some distributions, like the GEV, fit the right end tail better, while others have a better fit over the entire range of data. Our benchmark choice is the Weibull distribution because it maintains a better fit over the entire range of data for all severity classifications.

\medskip
\textbf{Types of financial crises}\medskip

Consistent with studies like \citet{ReinhartRogoff2009}, we find that currency crises lead to smaller output losses than debt and banking crises.\footnote{Distributions of these three types of crises' losses, calculated as losses accrued over a maximum period of five years until the average growth rate of a ten year pre-crises period is recovered, are depicted in figure 6, while descriptive statistics of all calculation methods are provided in tables 8 to 14.}
About 70 of the 122 banking crises in our sample lead to output losses. We can not conclude that banking and currency crises are generally preceded by high or low periods of growth as we do not observe a general dominance of losses calculated against three year pre-crisis trends as opposed to losses against ten year pre-crisis trends.

We find that average output losses after debt crises\footnote{23\% of initial GDP if recovery of the average HP \emph{growth rate} ten years before a crisis is considered (HP10 perc)} are 9\% higher than losses after banking crises. The median debt crisis is accompanied by output losses of 11.7\%, three percentage points larger than the median banking crisis. A large share of debt crises has, however, been accompanied by banking crises. Of the 62 debt crises in our sample, 36 have led to periods of negative growth. Of these 36 episodes, 26 have been accompanied by banking crises with mean losses about 30\% higher than if debt crises occur alone. Currency crises\footnote{According to HP10 perc} incur smaller losses than banking or debt crises of 15\% in the mean and 5\% in the median. This result is in line with \citet{Furcerietal2011} that concluded that debt crises tend to be more detrimental than banking and currency crises.

\begin{figure}[h!]
\centering
\caption{Severity distributions}\label{graph5}
\includegraphics[scale=0.4]{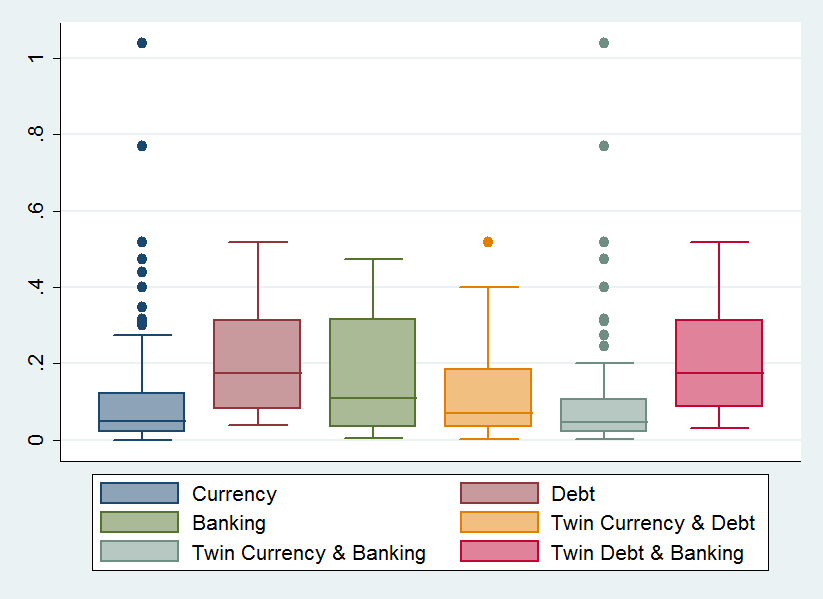}\\
\footnotesize{Source: Authors' calculations. For graphical reasons, distributions are cut at 1.5}
\end{figure}

Banking and debt crises alone are more severe than twin crises consisting of banking and currency crises or debt and currency crises. While twin crises between banking or debt and currency crises mostly lead to larger growth reductions in the very short run than banking or debt crises alone, long term losses are often found to be smaller. A possible explanation could be that a depreciation in the wake of a twin crisis (including a currency crisis) allows for a competitiveness gain which is not present when a banking crisis occurs alone and allows for a faster recovery. Twin crises consisting of debt and banking crises consequently incur the highest losses in our sample.\footnote{For detailed descriptive statistics of other loss measures, please see tables 8 to 14.}
\medskip

\textbf{Severity by region} \medskip

The highest losses after financial crises are experienced in Asia with average losses ranging from 9.8\% to 21.8\% of initial GDP.\footnote{Calculated against counterfactuals based on average HP filtered growth rates} The second highest losses are observed in Europe, followed by Latin America and Africa. The highest frequency of financial crises is, however, observed in Africa. Europe and Asia are mostly struck by currency and banking crises, while Africa and Latin America have suffered from all three types of crises to nearly equal degrees. In a joint analysis, debt crises were observed to be more severe than currency or banking crises. This result holds for all regions except for Asia, where banking crises lead to the most severe losses. Since a high share of currency and banking crises reported are in effect twin crises, it is however difficult to disentangle losses from both types separately. \citet{Furcerietal2011} confirm our results by controlling for the other types of crises.

\textbf{Severity by income groups}\medskip

In order to compare losses across different income groups, we classify countries into low, middle, and high income categories.\footnote{World Bank classification.} On average, middle income countries experience the highest output losses (15\%), followed by high and low income countries (12\% and 11\% of initial GDP respectively, see Table \ref{table3}). This observation holds for currency and banking crises, while losses from debt crises are almost exclusively observed in low and middle income countries.

The median crisis is more severe in high income than middle income countries. High average losses in middle income countries are driven by some extreme crisis events, such as Thailand in 1998, experiencing losses of 229\%.\footnote{See Figure \ref{graph7} in \ref{append1}.}  The median loss is 9\% for high income countries, 7\% for middle income countries, and 4\% for low income countries. In total, middle income countries suffer more from financial crises than high income countries as they experience a larger number of crises per country over the period observed.

\textbf{Severity of financial crises over time}\medskip

As expected, financial crises were especially harmful during the 1990's and depict the lowest losses during the 1970's and the period after the year 2000. However, this does not necessarily mean that financial crises have become smaller in magnitude.

\textbf{The severity of the 2007 financial crisis in the USA}
\medskip

Our main analysis considers a sample ending in 2005. However, we have enough information to estimate the severity of the 2007 financial crisis in the USA. We date the US banking crisis to late 2007 even though the main event, Lehman Brothers filing for Chapter 11 bankruptcy protection, did not take place until September 2008. Output trends are calculated as before with 2007 being the first year of output gap calculations. Data are taken from the IMF World Economic Outlook database until 2011. From 2012 until 2016, real GDP forecasts reported in the IMF database are used. Reliable output losses can therefore only be calculated over an interval of 5 years (2007-2011), larger loss windows are however reported for comparative reasons.

Figure \ref{graph10} in \ref{append1} shows real GDP for the USA from 1995 until 2016. Average real GDP growth in the ten years prior the financial crisis was 3.2 percent and 3.1 percent in the last three years before 2007. Real GDP growth slowed down to less than 2 percent in 2007 and finally turned negative during 2008 and 2009.

If the costs from the financial crisis are to be considered until the pre-crisis level of real GDP has been reached again and opportunity costs are ignored, the endpoint of the crisis is found to be in 2011 and costs remain modest at 5 percent of 2007 real GDP.

Turning towards our baseline calculations, calculated against a linear trend based on a pre-crisis time window of a HP filtered real GDP series, costs become larger and range between 15 and 48 percent of initial GDP if a 5 year time is considered and between 15 and 107 percent if costs are allowed to accumulate for up to ten years. The windows vary from a three year period, if the crisis is considered to be overcome once the pre-crisis average growth rate is recovered, to ten years in the case of defining the end of a crisis as recovery of the pre-crisis growth trend.

Baseline results over 5 year windows of around 30 percent of GDP are consistent with costs calculated in e.g. \citet{ChinnFrieden2011}, who estimate costs of the 2011 financial crisis at 3.5 trillion USD, which is roughly 26 percent of 2007 real GDP.

Historically, the 2007 financial crisis ranks among the most severe crises episodes. A comparable percentage loss of real GDP during the last 40 years has only been experienced by some Asian countries in the wake of the 1998 Asian crisis.

\subsection{The distribution of total losses}

Given our choice of the frequency distribution and the severity probability density function, we perform Monte Carlo simulations to obtain a numerical probability density function (PDF) of total losses over five years. These PDFs are markedly skewed to the right. Table \ref{table5} summarizes the results. The 99.9 percentile of the total loss distribution ranges from 2.4 to 3.6 trillion of 2005 USD. Estimates of cumulative world GDP loss due to the 2008-2009 world financial crisis are 5 trillion \citep{IMF2009}. This means that the recent financial crisis losses occur with very low probability. For the USA, \citet{ChinnFrieden2011} estimate a cumulative GDP loss for the USA to be about 3.5 trillions in 2005 USD.

%

\section{Risk-sharing against financial crises}

Can output losses from financial crises be diminished or crises prevented ex ante through an insurance scheme against rollover risk or capital flight? As financial crises are relatively rare events, one could imagine countries paying a certain amount of premium during financially stable times and in return having access to these funds during times of need.

Papers like \citet{Caballero2003} and \citet{Cordella2005,Cordella2006} have proposed insurance schemes with global scope. The estimation of the stochastic properties of financial crisis losses at the global perspective in terms of frequency, severity and their global aggregation is a useful device in thinking about the proposed insurance schemes.

According to our estimates, the potential worldwide costs from financial crises over periods of 5 years in percentage terms of 2005 world GDP are presented in Table \ref{table6}. Average costs of financial crises during a period a five years are relatively small and amount to less than one percent of 2005 world GDP. A period of extreme crisis events, occurring with a one percent probability, produces output costs of up to 4.54\% of World GDP. For example, if there is a will to cover global losses up to the 99 percentile, the amount of insurance coverage is such percentile minus the median value.

Many of the debt crises included in the above calculations are in fact destabilizing confidence crises or liquidity crises, during which rollover costs of debt become excessively high, rendering an illiquid country insolvent. In a similar manner, liquidity risk is often the cause for the occurrence of banking crises. The IMF provides a de facto insurance \citep[see]{Cordella2006} for these cases in the form of standard IMF programs. While the IMF does play the role of lender of last resort, it does so ex-post, after the country usually has already entered into financial turmoil.

As a type of national insurance against crises, nearly all emerging economies have accumulated large amounts of international reserves in order to possess a buffer against pro-cyclical international capital flows. \citet{Caballero2003} argues that hedging the financial mechanism behind macroeconomic disasters is a problem of a magnitude larger than a single market can handle.

The moral hazard problem arising through a potential insurance scheme is addressed in various studies. \citet{Cordella2005} examines to what degree the presence of a country insurance scheme affects the policymakers' incentives to undertake reforms. An important channel through which insurance can foster reforms can be identified: Insurance reduces the probability that deteriorating fundamentals evolve into large crises, which may enhance the expected political reforms and increase reform incentives.

Participation in a potential crisis insurance fund would therefore have to be subject to ex ante compliance with a number of clearly defined eligibility criteria, such as low budget deficits and a clear debt to GDP threshold. The potential insurance coverage must be forfeited as soon as the country does not fulfill all criteria. As stated in \citet{Cordella2006}, it would also be crucial to characterize and standardize the procedures followed after funds from the insurance facility have been accessed. In the optimal case, the existence of this insurance would incite fiscal discipline and, at the same time, provide liquidity if needed which in turn would lead to fewer crises.

Attractiveness of interest rate insurance implies that the insurance premium in tranquil times is lower than the costs through high rollover costs in turbulent times. In addition, the countries not experiencing liquidity problems have to be assured a benefit from the other country not entering into a crisis which exceeds the costs incurred through the insurance premium. Contrary to regular insurance, risk sharing depends on the contagion effects of financial crises from the country in financial turmoil to the liquid countries. Feasibility and benefit analysis is a potentially attractive area for future research.

\section{Conclusion}

Through the use of the financial crises database of \citet{Laeveneta2008} to date financial crises, we characterize the heterogeneity of aggregate output losses. In our LDA setup, the number of crisis events over a certain period is called the frequency of events, while the different metrics of GDP losses describe the severity of such losses. We fit common parametric frequency and severity distributions to compound global GDP loss densities and to report standard risk measures.

In line with the existing literature, we find that output losses after financial crises are strongly heterogeneous and a large number of countries never recover their pre-crises growth rates or trends. Loss distributions are skewed to the right, with average losses ranging between 9\% and 15\% of initial GDP.

Currency crises lead to smaller output losses than debt and banking crises, while the largest losses are found after debt crises. The presence of a debt crisis also exacerbates any of the other two forms of crises, while the presence of a currency crisis in the wake of a debt or banking crisis diminishes output losses through faster recovery. Banking and debt crises alone are found to be more severe than twin crises consisting of banking and currency crises or debt and currency crises.

We compare output costs from financial crises over world regions, country-income groups, and time. We find that Asia has suffered from the most severe financial crises, while Africa experiences the highest frequency of financial crises. Congruently, middle income countries experience the highest output losses, followed by high and low income countries. Financial crises are observed to have been especially harmful during the 1990's, while a global assessment of the severity of the recent 2008 financial crisis cannot yet be undertaken with our approach.

The LDA approach leads us to conclude that mean worldwide costs of financial crises within periods of 5 years are in the range of 0.5\% to 0.7\% of 2005 world GDP. Extreme crises episodes, occurring with a probability of one percent, can lead to losses in the range of 2.95\% to 4.54\% of world GDP.

There are some aspects that go beyond the scope of this paper. For example, the treatment of the appropriate time span for the LDA planning. We consider a five-year horizon but this is not necessarily the best from an optimal insurance design perspective. Also, all the calculations are based on aggregating GDP levels in comparable Dollar units. All calculations regarding percentage GDP losses are made after the LDA aggregation and relative to an initial GDP level. There is a possibility of exploring the LDA benchmark by allowing for the aggregation of percentage GDP losses directly. In this case, suitable GDP weights have to be considered. In addition, financial crisis distort long run growth in a number of cases, if so, losses in terms of GDP per capita instead of GDP levels can provide an adequate long run picture. Last, our loss approach is unconditional but the framework can be extended to incorporate conditioning factors in the stochastic processes that drive frequency and severity of GDP losses. We leave this issues open for future research.

\newpage
\small
\singlespace

\newpage

\pagebreak
\appendix
\renewcommand\thesection{Appendix \Alph{section}}

\renewcommand{\thefigure}{A-\arabic{figure}}
\setcounter{figure}{0}  

\section{Figures}\label{append1}

%
%

\begin{figure}[h!]
\centering
\includegraphics[scale=0.35]{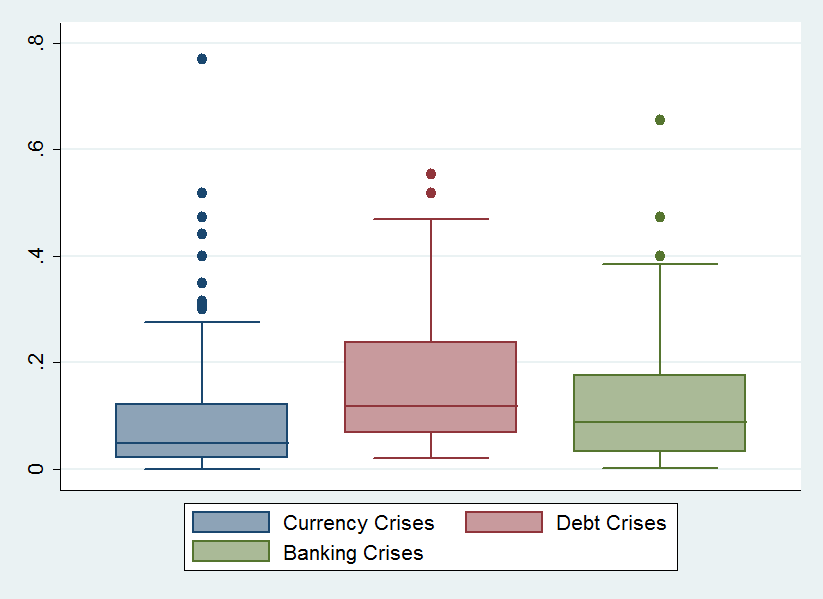}\\
\caption{Percentage loss distributions. Currency, debt, and banking crises}\label{graph6}
\end{figure}

\begin{figure}[h!]
\centering
\includegraphics[scale=0.35]{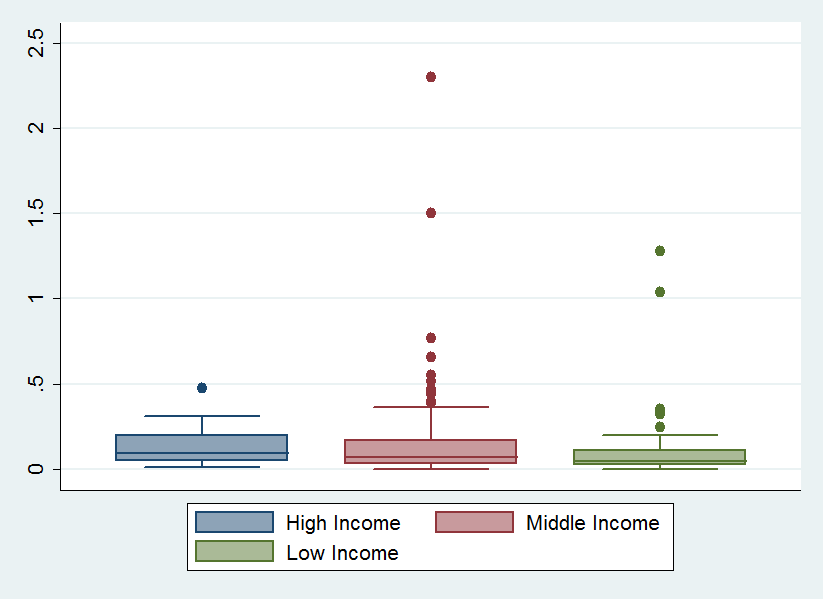}\\
\caption{Percentage loss distributions over income groups}\label{graph7}
\end{figure}

\newpage
\begin{figure}[h!]
\centering
\includegraphics[scale=0.35]{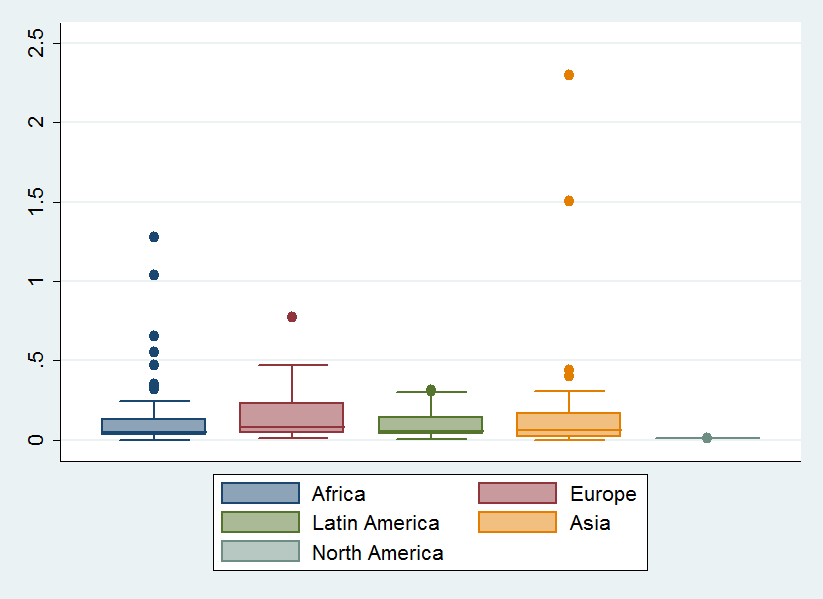}\\
\caption{Percentage loss distributions over regions}\label{graph8}
\end{figure}

\begin{figure}[h!]
\centering
\includegraphics[scale=0.35]{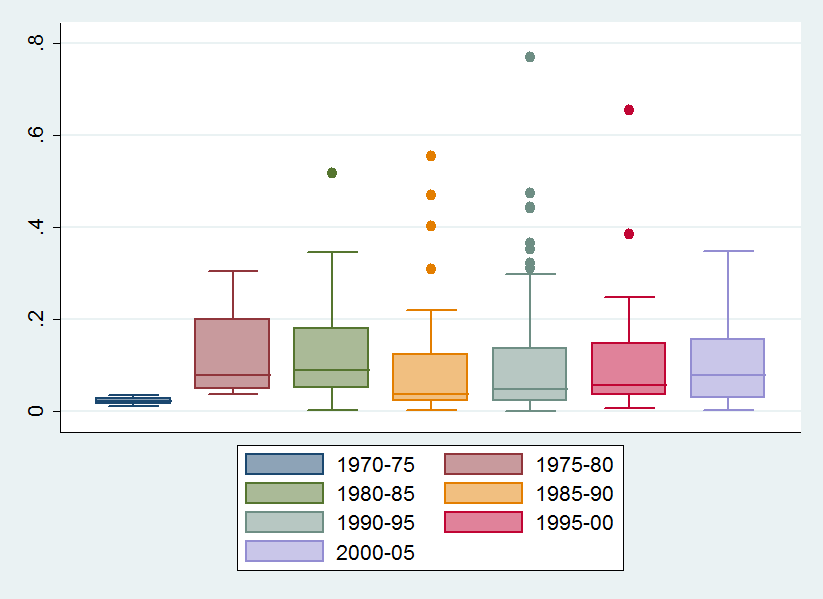}\\
\caption{Percentage loss distributions over time}\label{graph9}
\end{figure}

\newpage

\begin{figure}[h!]
\centering
\includegraphics[scale=0.7]{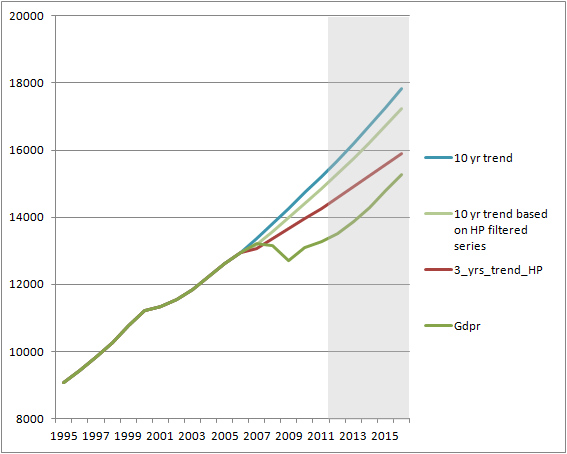}
\caption{Impact of the 2007 financial crisis on US growth}\label{graph10}
\end{figure}

\newpage

\section{Tables}\label{append2}

\begin{table}[!htb]
\centering
\caption{Severity of financial crises (as percentage of initial GDP)}\label{table1}
\begin{tabular} {l l l l l l l}
\hline
Loss Measure & Obs.     & Mean   & Median   & Std. Dev. & Min     & Max\\
\hline
HP10 perc    & 204      & 0.14  & 0.06    & 0.24     &0.0001   &2.30		 \\
HP10 trend   & 214	    & 0.15  & 0.07    & 0.25     &0.0003   &2.30 \\
HP3  perc    & 203      & 0.09  & 0.05    & 0.17     &0.0000   &2.04\\
HP3  trend   & 219      & 0.10  & 0.06    & 0.12     &0.0000   &1.06\\
\hline
\end{tabular}\\
\footnotesize{NOTE: Estimation of losses to recover average 10 year HP filtered GDP growth rates (HP10 perc), to recover the level of a 10 year HP filtered GDP trend (HP10 trend), to recover 3 year HP filtered GDP growth rates (HP3 perc), and to recover the level of a 3 year HP filtered GDP trend (HP3 trend)}
\end{table}

\begin{table}[!htb]
\centering
\caption{Severity of crises by Regions - All crises (Losses as percentage of initial GDP)}\label{table2}
\footnotesize{
\begin{tabular} {l l l l l l l l l l l}
\hline
Variable    &  Nr.      &Africa     &  Nr.  &Europe     &  Nr.  & Latin A.  &  Nr.  &Asia       &  Nr.  &North A.\\
\hline
HP10 perc  & 79        &0.13     &22     &0.17     &39     &   0.10  & 39    &  0.20   & 1     & 0.013\\
HP10 trend & 85        &0.13     &24     &0.18     &40     &   0.12  & 40    &  0.22   & 1     & 0.013\\
HP3 perc   & 78        &0.07     &24     &0.10     &41     &   0.12  & 36    &  0.10   & 1     & 0.008\\
HP3 trend  & 86        &0.08     &24     &0.14     &45     &   0.08  & 39    &  0.12   & 1     & 0.008\\
\hline
\end{tabular}\\
\footnotesize{NOTE: Estimation of losses to recover a 10 year HP filtered GDP growth  (HP10 perc), to recover the level of the 10 year HP filtered GDP (HP10 trend), to recover a 3 year HP filtered GDP growth  (HP3 perc) and to recover the level of the 3 year HP filtered GDP (HP3 trend)}
}
\end{table}

\begin{table}[!htb]
\centering
\caption{Severity by income groups (as percentage of initial GDP)}\label{table3}
\footnotesize{
\begin{tabular} {l l l l l l l}
\hline
Variable    &  Nr.  &High Income &  Nr. &Middle Income  &  Nr.          & Low  Income\\
\hline
HP10 perc  &21     &  0.13    &122    & 0.18       &61             &   0.11\\
HP10 trend &22     &  0.16    &128    & 0.17       &64             &   0.12\\
HP3 perc   &22     &  0.07    &120    & 0.11       &61             &   0.07\\
HP3 trend  &23     &  0.09    &128    & 0.11       &68             &   0.07\\
\hline

\end{tabular}
\\
\footnotesize{NOTE: Estimation of losses to recover a 10 year HP filtered GDP growth  (HP10 perc), to recover the level of the 10 year HP filtered GDP (HP10 trend), to recover a 3 year HP filtered GDP growth  (HP3 perc) and to recover the level of the 3 year HP filtered GDP (HP3 trend)}
}
\end{table}

\begin{table}[!htb]
\centering
\caption{Severity over time - All crises} (as percentage of initial GDP)\label{table4}
\begin{tabular} {l l l l l}
\hline
Period      & HP10 perc & HP10trend & HP3 perc & HP3 trend \\
\hline
1970 -75 & 0.02 & 0.05 & 0.02 & 0.06 \\
1975 -80 & 0.11 & 0.14 & 0.19 & 0.09 \\
1980 -85 & 0.18 & 0.14 & 0.09 & 0.10 \\
1985 -90 & 0.12 & 0.10 & 0.06 & 0.07 \\
1990 -95 & 0.15 & 0.16 & 0.09 & 0.12 \\
1995 -00 & 0.19 & 0.20 & 0.06 & 0.07 \\
2000 -05 & 0.11 & 0.12 & 0.07 & 0.07 \\
\hline
\end{tabular}
\\
\footnotesize{NOTE: Estimation of losses to recover a 10 year HP filtered GDP growth  (HP10 perc), to recover the level of the 10 year HP filtered GDP (HP10 trend), to recover a 3 year HP filtered GDP growth  (HP3 perc) and to recover the level of the 3 year HP filtered GDP (HP3 trend)}

\end{table}

\medskip

\begin{table}[!htb]
\centering
\caption{Features of the PDF of total losses.}\label{table5}
\vspace{0.2cm}
\begin{tabular} {lcccc}
\hline
                 & HP10perc          & HP10trend          & HP3perc          & HP3trend\\
\hline
{99.9 percentile} &3.0e+012           &3.6e+012            &2.4e+012          &2.7e+012\\
{99 percentile}   &1.7e+012           &2.0e+012            &1.3e+012          &1.5e+012\\
{median}          &1.9e+011           &2.2e+011            &1.4e+011          &1.6e+011\\
{mean}            &3.0e+011           &3.6e+011            &2.3e+011          &2.7e+011\\
{Std deviation}   &3.7e+011           &4.4e+011            &2.9e+011          &3.3e+011\\
\hline
\multicolumn{5}{l}{\footnotesize{Notes:}}\\
\multicolumn{5}{l}{\footnotesize{(1) Number of simulations equal to 500 000. In all cases, the frequency distribution}}\\
\multicolumn{5}{l}{\footnotesize{\hspace{17pt} is the Negative Binomial, the severity PDF is a Weibull}}\\
\multicolumn{5}{l}{\footnotesize{(2) Losses are measured in constant 2005 USD and correspond to five-year periods}}\\
\end{tabular}\\
\vspace{0.2cm}
\end{table}

\begin{table}[!htb]
\centering
\caption{LDA, Losses in constant 2005 USD, over periods of five years, as percentage of 2005 World GDP}\label{table6}
\vspace{0.2cm}
\begin{tabular} {l l l l l}
\hline
                 & HP10perc      & HP10trend    & HP3perc   & HP3trend              \\
\hline
99.9\% perctl    & $6.81\%  $     & $8.18\%$      & $5.45\%$   & $6.13\%$ \\
99 \% perctl     & $3.86\%  $     & $4.54\%$      & $2.95\%$   & $3.40\%$ \\
Median           & $0.43\%$       & $0.50\%$      & $0.31\%$   & $0.36\%$ \\
Mean             & $0.68\%$       & $0.81\%$      & $0.52\%$   & $0.61\%$ \\
\hline
\multicolumn{5}{l}{\footnotesize{Notes:}}\\
\multicolumn{5}{l}{\footnotesize{(1) Number of simulations equal to 500 000. In all cases, the frequency distribution}}\\
\multicolumn{5}{l}{\footnotesize{\hspace{17pt} is the Negative Binomial, the severity pdf is a Weibull, 2005 World GDP from WDI2006}}\\
\end{tabular}\\
\vspace{0.2cm}
\end{table}

\begin{table}[!htb]
\centering
\caption{Losses after financial crises (as percentage of initial GDP)}\label{table7}
\begin{tabular} {l l l l l l l}
\hline
Loss Measure   & Obs.     & Mean   & Median   & Std. Dev. & Min & Max\\
\hline
AG(10)(5)perc  & 186      & 0.3241 & 0.1632   &0.4358     &0.0005   &2.7310   \\
AG(10)(5)trend & 186	  & 0.4257 & 0.2417   &0.5217     &0.0005   &3.6159		 \\
AG(3)(5)perc   & 182	  & 0.3249 & 0.1696   &0.3997     &0.0000   &2.0615	 \\
AG(3)(5)trend  & 186      & 0.4567 & 0.2776   &0.5667     &0.0000   &4.4838  \\
ABS            & 110      & 0.8509 & 0.1111   &2.2077     &0.0000   &16.0377\\
AG(10)(10)perc & 186      & 0.5441 & 0.1632   &1.1689     &0.0005   &8.2695\\
AG(10)(10)trend& 186      & 1.0587 & 0.2604   &1.9302     &0.0005   &11.1335\\
AG(3)(10)perc  & 180      & 0.5809 & 0.1732   &1.2679     &0.0000   &9.2046\\
AG(3)(10)trend & 181      & 0.9878 & 0.3403   &1.6139     &0.0000   &9.2046\\
\hline
\end{tabular}
\end{table}

\medskip
\vspace{2cm}

\begin{table}[!ht]
\centering
\caption{Losses after currency crises (as percentage of initial GDP)}\label{table8}
\begin{tabular} {l l l l l l}
\hline
Variable              & Obs    & Mean      & Std. Dev.  & Min     & Max\\
\hline
AG(10)(5)perc         & 101    &0.2794     &0.3949      &0.0006   &1.8448\\
AG(10)(5)trend        & 101    &0.3702     &0.4870      &0.0006   &2.2125\\
AG(3)(5)perc          & 94     &0.2861     & 0.3658     &0.0000   &1.7087\\
AG(3)(5)trend         & 94     &0.3874     &0.4260      &0.0000   &1.7598\\
HP(10)perc            & 118    &0.1352     &0.2825      &0.0001   &2.2993\\
HP(10)trend           & 122    &0.1563     &0.2947      &0.0003   &2.2993\\
HP(3)perc             & 114    &0.0918     &0.2111      &0.0013   &2.0370\\
HP(3)trend            & 125    &0.0904     &0.1293      &0.0013   &1.0627\\
ABS                   & 61     &0.6811     &1.4726      &0.0001   &7.0657\\
AG(10)(10)perc        & 101    &0.3674     &0.7658      &0.0006   &6.2588\\
AG(10)(10)trend       & 101    &0.8587     &1.8207      &0.0006   &11.1335\\
AG(3)(10)perc         & 94     &0.3635     &0.7025      &0.0000   &5.4637\\
AG(3)(10)trend        & 94     &0.8509     &1.4290      &0.0000   &8.3445\\
\hline
\end{tabular}
\end{table}

\medskip
\vspace{2cm}

\begin{table}[!htb]
\centering
\caption{Losses after banking crises (as percentage of initial GDP)}\label{table9}
\begin{tabular} {l l l l l l}
\hline
Variable              & Obs   & Mean    & Std. Dev. & Min     & Max\\
\hline
AG(10)(5)perc         & 69    &0.3650   &0.4636     &0.0033   &2.7310  \\
AG(10)(5)trend        & 69    &0.4834   &0.5726     &0.0033   &3.6159\\
AG(3)(5)perc          & 71    &0.3321   &0.3925     &0.0049   &2.0615\\
AG(3)(5)trend         & 71    &0.4793   &0.6519     &0.0049   &4.4838\\
HP(10)perc            & 66    &0.1424   &0.1914     &0.0014   &1.2766\\
HP(10)trend           & 71    &0.1547   &0.2084     &0.0014   &1.3075\\
HP(3)perc             & 70    &0.0800   &0.0763     &0.0000   &0.4705\\
HP(3)trend            & 74    &0.0958   &0.1075     &0.0000   &0.7574\\
ABS                   & 35    &0.7631   &2.1609     &0.0000   &12.5942\\
AG(10)(10)perc        & 69    &0.7637   &1.5744     &0.0033   &8.2695\\
AG(10)(10)trend       & 69    &1.5239   &2.4667     &0.0033   &11.1335\\
AG(3)(10)perc         & 69    &0.7888   &1.7237     &0.0049   &9.2046\\
AG(3)(10)trend        & 70    &1.7113   &4.7780     &0.0049   &9.2046\\
\hline
\end{tabular}
\end{table}

\newpage

\begin{table}[!htb]
\centering
\caption{Losses after debt crises (as percentage of initial GDP)}\label{table10}
\begin{tabular} {l l l l l l}
\hline
Variable   & Obs    & Mean     & Std. Dev. & Min       & Max\\
\hline
AG(10)(5)perc & 39     &0.5002    &0.4842     &0.0005     &1.6946 \\
AG(10)(5)trend & 39     &0.6244    &0.4797     &0.0005     &1.6946\\
AG(3)(5)perc  & 39     &0.4282    &0.4024     &0.0005     &1.8020\\
AG(3)(5)trend  & 39     &0.5248    &0.4477     &0.0005     &1.8020\\
HP(10)perc     & 46     &0.1723    &0.1413     &0.0211     &0.5531\\
HP(10)trend  & 47     &0.1966    &0.1597     &0.0058     &0.7095\\
HP(3)perc   & 43     &0.1224    &0.1079     &0.0089     &0.4846\\
HP(3)trend   & 45     &0.1395    &0.1108     &0.0025     &0.4846\\
ABS    & 36     &0.8658    &2.7270     &0.0026     &16.0377\\
AG(10)(10)perc  & 39     &1.6642    &0.8180     &0.0005     &4.0388\\
AG(10)(10)trend   & 39     &1.8217    &2.8550     &0.0005     &11.1335\\
AG(3)(10)perc   & 39     &0.6273    &0.9522     &0.0005     &4.6050\\
AG(3)(10)trend & 39     & 1.1286   & 1.4862    &0.0005     &5.7084\\
\hline
\end{tabular}
\end{table}

\begin{table}[!htb]
\centering
\caption{Twin crises - Currency and Debt (as percentage of initial GDP)}\label{table11}
\begin{tabular} {l l l l l l}
\hline
Variable  & Obs   & Mean   & Std. Dev.  & Min    & Max\\
\hline
AG(10)(5)perc  & 56    &0.3983  &0.4209  &0.0005      &1.8448\\
AG(10)(5)trend  & 56    &0.5286  &0.5121  &0.0005      &2.2125\\
AG(3)(5)perc  & 52    &0.3849  &0.4200  &0.0005      &1.8020\\
AG(3)(5)trend   & 52    &0.5198  &0.4797  &0.0005      &1.8020\\
HP(10)perc  & 64    &0.1334  &0.1223  &0.0018      &0.5170\\
HP(10)trend   & 67    &0.1561  &0.1520  &0.0018      &0.7095\\
HP(3)perc  & 66    &0.1172  &0.2541  &0.0023      &2.0370\\
HP(3)trend  & 71    &0.1032  &0.0998  &0.0023      &0.4846\\
ABS     & 42    &0.5397  &1.2529  &0.0026      &7.0657\\
AG(10)(10)perc & 56    &0.5708  &0.9698  &0.0005      &6.2588\\
AG(10)(10)trend  & 56    &1.4029  &2.5465  &0.0005      &11.1335\\
AG(3)(10)perc  & 52    &0.5311  &0.9354  &0.0005      &5.4637\\
AG(3)(10)trend & 52    &1.0760  &1.3452  &0.0005      &4.7534\\
\hline
\end{tabular}
\end{table}

\begin{table}[!htb]
\centering
\caption{Twin crises - Currency and Banking (as percentage of initial GDP)}\label{table12}
\begin{tabular} {l l l l l l}
\hline
Variable  & Obs   & Mean     & Std. Dev.  & Min        & Max\\
\hline
AG(10)(5)perc   & 78    &0.2716    &0.3726      &0.0005      &1.6946\\
AG(10)(5)trend  & 78    &0.3435    &0.4417      &0.0005      &1.6946\\
AG(3)(5)perc & 73    &0.2965    &0.3360      &0.0005      &1.8020\\
AG(3)(5)trend  & 73    &0.3900    &0.3884      &0.0005      &1.8020\\
HP(10)perc & 101   &0.1531    &0.3014      &0.0018      &2.2993\\
HP(10)trend  & 103   &0.1716    &0.3123      &0.0018      &2.2993\\
HP(3)perc  & 95    &0.0822    &0.1108      &0.0013      &0.8544\\
HP(3)trend  & 103   &0.0960    &0.1342      &0.0013      &1.0627\\
ABS      & 56    &0.6501    &1.3273      &0.0001      &6.8260\\
AG(10)(10)perc  & 78    &0.3051    &0.4492      &0.0005      &2.1887\\
AG(10)(10)trend & 78    &0.8651    &2.1069      & 0.0005     &11.1335\\
AG(3)(10)perc   & 73    &0.3492    &0.5440      &0.0005      &3.7145\\
AG(3)(10)trend & 73    &0.7216    &1.1281      &0.0005      &5.7125\\
\hline
\end{tabular}
\end{table}

\newpage

\begin{table}[!htb]
\centering
\caption{Twin crises - Debt and Banking (as percentage of initial GDP)}\label{table13}
\begin{tabular} {l l l l l l}
\hline
Variable & Obs  & Mean   & Std. Dev.  & Min        & Max\\
\hline
AG(10)(5)perc   & 36   &0.5213  &0.5045      &0.0005      &1.6946\\
AG(10)(5)trend  & 36   &0.6631  &0.4996      &0.0005      &1.6946\\
AG(3)(5)perc & 35   &0.4247  &0.3862      &0.0005      &1.8020\\
AG(3)(5)trend  & 35   &0.5553  &0.4129      &0.0005      &1.8020\\
HP(10)perc  & 32   &0.2396  &0.2447      &0.0093      &1.2766\\
HP(10)trend  & 34   &0.2586  &0.2595      &0.0058      &1.3075\\
HP(3)perc   & 32   &0.1510  &0.1293      &0.0000      &0.4846\\
HP(3)trend  & 33   &0.1782  &0.1590      &0.0000      &0.7574\\
ABS    & 26   &1.1814  &3.1728      &0.0059      &16.0377\\
AG(10)(10)perc & 36   &0.7952  &1.1590      &0.0005      &5.4650\\
AG(10)(10)trend  & 36   &1.7638  &2.6372      &0.0005      &11.1335\\
AG(3)(10)perc  & 34   &0.8204  &1.7036      &0.0005      &9.2046\\
AG(3)(10)trend  & 35   &1.3295  &1.9641      &0.0005      &9.2046\\
\hline
\end{tabular}
\end{table}

\begin{table}[!htb]
\centering
\caption{All Twin Crises (as percentage of initial GDP)}\label{table14}
\begin{tabular} {l l l l l l}
\hline
Variable    & Obs  & Mean     & Std. Dev. & Min    & Max\\
\hline
AG(10)(5)perc    & 114  &  0.3173  &  0.4052 &  0.0005  &    1.8448\\
AG(10)(5)trend  & 114  &  0.4238  &  0.4946 &  0.0005  &    2.2125\\
AG(3)(5)perc  & 105  &  0.3344  &  0.3755 &  0.0005  &    1.8020\\
AG(3)(5)trend  & 105  &  0.4527  &  0.4386 &  0.0005  &    1.8020\\
HP(10)perc  & 130  &  0.1588  &  0.2902 &  0.0018  &    2.2993\\
HP(10)trend  & 136  &  0.1759  &  0.2996 &  0.0018  &    2.2993\\
HP(3)perc  & 126  &  0.1068  &  0.2071 &  0.0000  &    2.0370\\
HP(3)trend & 136  &  0.1078  &  0.1423 &  0.0000  &    1.0627\\
ABS      & 78   &  0.9035  &  2.2318 &  0.0001  &    16.0377\\
AG(10)(10)perc  & 114  &  0.4741  &  0.9509 &  0.0005  &    6.2588\\
AG(10)(10)trend  & 114  &  0.9907  &  2.0026 &  0.0005  &    11.1335\\
AG(3)(10)perc   & 104  &  0.5286  &  1.1851 &  0.0005  &    9.2046\\
AG(3)(10)trend  & 105  &  0.9816  &  1.5407 &  0.0005  &    9.2046\\
\hline
\end{tabular}
\centering
\end{table}

\begin{table}[!htb]
\centering
\footnotesize{
\caption{Severity of crises by regions - All crises (Losses as percentage of initial GDP)}\label{table15}
\begin{tabular} {l l l l l l l l l l l}
\hline
Variable       &  Nr.      &Africa     &  Nr.  &Europe     &  Nr.  & Latin A.  &  Nr.  &Asia       &  Nr.  &North A.\\
\hline
AG(10)(5)perc  &  71       &0.2481     &21     & 0.4674    &37     &   0.2759  & 32    &  0.5112   & 1     & 0.0091\\
AG(10)(5)trend &  71       &0.3253     &21     & 0.6047    &37     &   0.3909  & 32    &  0.5751   & 1     & 0.0091\\
AG(3)(5)perc   & 62        &0.2309     &26     & 0.4075    &35     &   0.3083  & 32    &  0.5131   & 1     & 0.0390\\
AG(3)(5)trend  & 62        &0.3080     &26     & 0.5250    &35     &   0.4170  & 32    &  0.6949   & 1     & 0.0390\\
ABS            & 43        &0.7816     &17     &1.2325     &21     &   0.3113  & 14    &  1.4106   & 0     & 0     \\
AG(10)(10)perc & 71        &0.4279     &21     &0.5354     &37     &   0.3072  & 32    &  1.0558   & 1     & 0.0091\\
AG(10)(10)trend& 71        &0.9220     &21     &1.0626     &37     &   0.7603  & 32    &  1.8413   & 1     & 0.0091\\
AG(3)(10)perc  & 60        &0.3427     &26     &0.5793     &35     &   0.3789  & 32    &  1.1557   & 1     & 0.0390\\
AG(3)(10)trend & 61        &0.6376     &26     &1.2732     &35     &   0.6525  & 32    &  1.6113   & 1     & 0.0390\\
\hline
\end{tabular}
}
\end{table}

\begin{table}[!htb]
\centering
\caption{Average severity of crises by income groups (percentage of initial GDP)}\label{table16}
\begin{tabular} {l l l l l l l}
\hline
Variable        &  Nr.  &High Income &  Nr. &Middle Income  &  Nr.          & Low  Income \\
\hline
AG(10)(5)perc   &21     &  0.2844    &112    & 0.3801       &53             &   0.2215\\
AG(10)(5)trend  &21     &  0.3901    &112    & 0.4945       &53             &   0.2946\\
AG(3)(5)perc    &21     &  0.2389    &113    & 0.3800       &48             &   0.2327\\
AG(3)(5)trend   &21     &  0.4923    &113    & 0.4915       &48             &   0.2890\\
ABS             &10     &  0.4446    &71     & 0.7927       &29             &   1.1332\\
AG(10)(10)perc  &21     &  0.4032    &112    & 0.6669       &53             &   0.3403\\
AG(10)(10)trend &21     &  1.0673    &112    & 1.1918       &53             &   0.7740\\
AG(3)(10)perc   &21     &  0.4258    &112    & 0.6814       &47             &   0.4106\\
AG(3)(10)trend  &21     &  1.1015    &113    & 1.0663       &47             &   0.7485\\
\hline

\end{tabular}

\end{table}

\newpage



\newpage

\begin{table}[!htb]
\centering
\footnotesize
\caption{Features of the PDF of total losses (more cases)}\label{table17}
\centering
\begin{tabular}{lccccc}
\hline
                & ABS     &AG10.10perc  &AG10.10trend &AG3.10perc  & AG3.10trend\\
\hline
99.9 percentile  &5.7e+012 &1.8e+013     &3.7e+013     &2.2e+013    &4.7e+013\\
99 percentile    &2.7e+012 &8.7e+012     &1.7e+013     &1.1e+013    &2.2e+013\\
median           &1.8e+011 &6.4e+011     &1.2e+012     &8.1e+011    &1.6e+012\\
mean             &3.8e+011 &1.3e+012     &2.4e+012     &1.6e+012    &3.1e+012\\
Std. deviation   &6.0e+011 &1.9e+012     &3.8e+012     &2.3e+012    &4.8e+012\\
\hline
\multicolumn{6}{l}{\footnotesize{Notes:}}\\
\multicolumn{6}{l}{\footnotesize{(1) Number of simulations equal to 500 000. In all cases, the frequency distribution}}\\
\multicolumn{6}{l}{\footnotesize{\hspace{17pt} is the Negative Binomial, the severity pdf is a Weibull}}\\
\multicolumn{6}{l}{\footnotesize{(2) Losses are measured in constant 2005 USD and correspond to five-year periods}}\\
\end{tabular}

\end{table}

\begin{table}[!htb]
\centering
\caption{Features of the PDF of total losses (more cases)}\label{table18}
\vspace{0.2cm}
\begin{tabular}{lcccc}
\hline
                & AG10.5perc &AG10.5trend & AG3.5perc &AG3.5trend\\\hline
99.9 percentile  &8.8e+012    &1.2e+013    &1.1e+013   &1.5e+013\\
99 percentile    &4.5e+012    &6.2e+012    &5.4e+012   &7.8e+012\\
median           &4.0e+011    &5.4e+011    &4.9e+011   &6.9e+011\\
mean             &7.3e+011    &9.8e+011    &8.6e+011   &1.2e+012\\
Std. deviation   &9.8e+011    &1.3e+012    &1.2e+012   &1.7e+012\\
\hline
\multicolumn{5}{l}{\footnotesize{Notes:}}\\
\multicolumn{5}{l}{\footnotesize{(1) Number of simulations equal to 500 000. In all cases, the frequency distribution}}\\
\multicolumn{5}{l}{\footnotesize{\hspace{17pt} is the Negative Binomial, the severity pdf is a Weibull}}\\
\multicolumn{5}{l}{\footnotesize{(2) Losses are measured in constant 2005 USD and correspond to five-year periods}}\\
\end{tabular}\\
\vspace{0.2cm}
\end{table}

\newpage

\begin{table}[!htb]
\centering
\caption{USA 2007 banking crisis losses (percentage of 2007 real GDP)}\label{table19}
\begin{tabular} {l r r}
\hline
Loss measure    & in Billion USD & \% of 2007 GDP  \\
\hline
HP(10)(10)perc   &3009.28         &  22.78    \\
HP(10)(10)trend  &14027.9         &  106.22   \\
HP(3) (10)perc   &2014.14         &  15.25    \\
HP(3) (10)trend  &7415.09         &  56.14    \\
HP(3) (5) trend  &4043.20         &  30.61    \\
HP(3) (5) perc   &2014.14         &  15.25    \\
HP(10)(5) trend  &6329.73         &  47.92    \\
HP(10)(5) perc   &3009.28         &  22.78    \\
AG(10)(5) perc   &5923.15         &  44.85     \\
AG(10)(5) trend  &5923.15         &  44.85     \\
AG(3) (5) perc   &5552.58         &  42.04     \\
AG(3) (5) trend  &5552.58         &  42.04     \\
AG(10)(10)perc   & 15334.18       & 116.11     \\
AG(10)(10)trend  & 17874.42       & 135.34     \\
AG(3) (10)perc   & 9700.10        & 73.45      \\
AG(3) (10)trend  & 17851.69       & 135.17     \\
ABS              &666.10          &  5.04      \\
\hline

\end{tabular}
\end{table}


\begin{thebibliography}{99}

\bibitem[Aziz et.al.(2000)]{Azizeta2000} Aziz, Jahangir, Francesco Caramazza and Ranil Salgado. (2000). ``Currency Crises: In Search of Common Elements'', IMF Working Papers 00/67.

\bibitem[Angkinand(2008)]{Angkinand2008} Angkinand, Apanard Penny. (2008). ``Output Loss and Recovery from Banking and Currency Crises:Estimation Issues'', manuscript.

\bibitem[Barro(2001)]{Barro2001} Barro, Robert. (2001). ``Economic Growth in East Asia before and after the Financial Crises'', NBER Working Paper 8330.

\bibitem[Barro(2006)]{Barro2006} Barro, Robert. (2006). ``Rare Disasters and Asset Markets in the Twentieth Century'', The Quarterly Journal of Economics 121(3), pp. 823-866.

\bibitem[Bicaba et.al.(2011)]{BKM2011} Bicaba, Zorobabel, Daniel Kapp and Franceso Molteni. (2011). ``Stability periods between Financial crises: The role of macroeconomic fundamentals and crises management policies'', CES Working Paper 2011.64


\bibitem[Bordo et.al.(2001)]{Bordoetal2001} Bordo, Michael, Barry Eichengreen, Daniela Klingebiel and Maria Soledad Martinez-Peria. (2001). ``Is the crisis problem growing more severe?'', Economic Policy, 16(32), pp. 51-82.

\bibitem[Borio et.al.(1994)]{Borioetal1994} Borio, Claudio, N. Kennedy and Stephen David Prowse. (1994). ``Exploring aggregate asset price fluctuations across countries: measurement, determinants and monetary policy implications'', BIS Economic Papers, 40, April.

\bibitem[Boyd et.al.(2000)]{Boydeta2000} Boyd, John, Pedro Gomis, Sungkyu Kwak and Bruce Smith. (2000). ``A User's Guide to Banking Crises'', manuscript.

\bibitem[Boyd et.al.(2002)]{Boydeta2005} Boyd, John, Sungkyu Kwak and Bruce Smith. (2005). ``The Real Output Losses Associated with Modern Banking Crises'', Journal of Money, Credit and Banking, 37(6), pp. 977-999.

\bibitem[Caballero(2003)]{Caballero2003} Caballero, Ricardo. (2003). ``The Future of the IMF'', American Economic Review, 93(2), pp. 31-38.



\bibitem[Caprio et.al.(1996)]{Caprioeta1996} Caprio, Gerard and  Daniela Klingebiel. (1996). ``Bank Insolvencies. Cross-country Experience'', Policy Research Working Paper 1620, The World Bank.




\bibitem[Cecchetti et.al.(2009)]{Cecchetteta2009} Cecchetti, Stephen, Marion Kholer and Christian Upper. (2009). ``Financial Crises and Economic Activity'', NBER Working Papers 15379.




\bibitem[Chernobai et.al.(2011)]{Chernobaietal2007} Chernobai, Anna, Svetlozar Rachev and Frank Fabozzi. (2011). Operational Risk: A Guide to Basel II Capital Requirements, Models, and Analysis. Wiley Finance.

\bibitem[Chinn and Frieden(2011)]{ChinnFrieden2011} Chinn, Menzie and Jeffry Frieden. (2011). Lost Decades: The Making of American Debt Crisis and Recovery. W.W. Norton \& Company, Inc.



\bibitem[Cordella and Levy-Yeyati(2005)]{Cordella2005} Cordella, T., Yeyati, E.L. (2005), ``Country Insurance'' , \textit{IMF Staff Paper}, IMF, Vol.52, Special Issue.

\bibitem[Cordella and Yeyati(2006)]{Cordella2006} Cordella, Tito, Eduardo Levy-Yeyati (2006), ``A (New) Country Insurance Facility'' International Finance, 9(1), pp. 1-36.

\bibitem[Cruz(2004)]{Cruz2004} Cruz, Marcelo. (2004). Operational risk modelling and analysis: theory and practice, London: Risk Books.


\bibitem[Dziobek et.al.(2008)]{Dziobeketa1997} Dziobek, Claudia and Ceyla Pazarbasioglu. (1997). ``Lessons from Systemic Bank Restructuring: A Survey of 24 Countries'', IMF Working Paper 97/161.

\bibitem[Demirguc-Kunt and Detragiache(1998)]{Demirguc-Kunt1998} Demirguc-Kunt, Asli and Enrica Detragiache. (1998). ``The Determinants of Banking Crises in Developing and Developed Countries'', \textit{IMF Staff Papers}, 45(1), pp. 81-109.

\bibitem[Demirguc-Kunt et.al.(2006)]{Demirguc-Kunt2006} Demirguc-Kunt, Asli, Enrica Detragiache and Poonam Gupta.(2006). ``Inside the crisis: An empirical analysis of banking systems in distress'', Journal of International Money and Finance, 25(5), 702-718.





\bibitem[Furceri et.al.(2011)]{Furcerietal2011} Furceri, Davide and Davide Zdzienicka. (2011). ``How Costly Are Debt Crises?'' , IMF Working Papers 11/280.

\bibitem[Frankel and Rose(1996)]{FrankelRose1996} Frankel, Jeffrey and Andrew Rose. (1996). ``Currency Crashes in Emerging Markets: An Empirical Treatment.'' Journal of International Economics 41, pp. 351-66.

\bibitem[Frydl(1999)]{Frydl1999} Frydl, Edward. (1999). ``The Length and Cost of Banking Crises'', IMF Working Papers 99/30.

\bibitem[Gorton(1988)]{Gorton1988} Gorton, Gary. (1988). ``Banking Panics and Business Cycles'', Oxford Economic Papers, 40 pp. 221-55.

\bibitem[Gupta et.al.(2007)]{Guptaeta2007} Gupta, Poonam, Deepak Mishra and Ratna Sahay. (2007). ``Behavior of output during currency crises'' , Journal of International Economics, 72, pp. 428-450.

\bibitem[Hanna and Huang(2002)]{HannaHuang2002} Hanna, Don and Yiping Huang. (2002). ``Bank Restructuring in Post-Crisis Asia'', Asian Economic Papers, 1(1), pp. 3-42.


\bibitem[Hoggarth et.al.(2002)]{Hoggarth2002} Hoggarth, Glenn, Ricardo Reis and Victoria Saporta. (2002). ``Costs of banking system instability: some empirical evidence'', Journal of Banking \& Finance, 26(5), pp. 825-855.



\bibitem[Hutchison(1999)]{Hutchison1999} Hutchison, •Michael and Kathleen McDill. (1999). ``Are All Banking Crises Alike? The Japanese Experience in International Comparison.'' Journal of the Japanese and International Economics, 13, pp. 155-180.



\bibitem[IMF(2009)]{IMF2009} IMF (2009), Global Financial Stability Report. November 2009 update.

\bibitem[Kaminsky and Reinhart(1999)]{KaminskyReinhart1999} Kaminsky, Graciela and Carmen Reinhart. (1999). ``The Twin Crises: The Causes of Banking and Balance-of-Payments Problems.'' American Economic Review, 89(3), pp. 473-500.


\bibitem[Kindleberger(1978)]{Kindleberger1978} Kindleberger, Charles. (1978). Manias, Panics, and Crashes: A History of Financial Crises, Palgrave Macmillan.

\bibitem[Laeven and Valencia(2008)]{Laeveneta2008} Laeven, Luc and Fabian Valencia, (2008). ``Systemic Banking Crises: A New Database'', IMF Working Paper 08/224.

\bibitem[Lindgren et.al.(1996)]{LindgrenGarciaSaal1996} Lindgren, Carl-Johan, Gillian Garcia and Matthew Saal. (1996). Bank Soundness and Macroeconomic Policy, Washington, D.C.: International Monetary Fund.

\bibitem[Logan(2001)]{Logan2001} Logan, Andrew. (2001). ``The United Kingdom's small banks' crisis of the early 1990s: what were the leading indicators of failure?'', Bank of England Working Paper 139.



\bibitem[Panjer(2006)]{Panger2006} Panjer, Harry. (2006). Operational Risk: Modeling Analytics. Wiley, New York




\bibitem[Reinhart and Rogoff(2009)]{ReinhartRogoff2009} Reinhart, Carmen and Kenneth Rogoff. (2009). This Time Is Different: Eight Centuries of Financial Folly, Princeton University Press.

\bibitem[Shevchenko(2011)]{Shevchenko2011} Shevchenko, Pavel. (2011). Modelling Operational Risk using Bayesian Inference, Springer, Berlin.
\end{thebibliography}
\end{document}